\documentclass[fleqn,usenatbib]{mnras}
\usepackage{newtxtext,newtxmath}

\usepackage{soul}

\usepackage[T1]{fontenc}
\usepackage{ae,aecompl}

\usepackage{graphicx}	
\usepackage{amsmath}	
\usepackage{amssymb}	

\title[OIII Variability in RZ2109]{Slow Decline and Rise of the Broad [OIII] Emission Line in Globular Cluster Black Hole Candidate RZ2109}

\author[K.C. Dage et al]{
Kristen C. Dage,$^{1}$\thanks{E-mail: kcdage@msu.edu}
Stephen E. Zepf,$^{1}$, 
Arash Bahramian, $^{1, 2}$
Jay Strader, $^{1}$
\newauthor
Thomas J. Maccarone, $^3$
Mark B. Peacock,$^{1}$ 
Arunav Kundu, $^4$
\newauthor
Matthew M. Steele $^1$
Christopher T. Britt  $^{1,5}$
\\
$^{1}$Department  of  Physics  and  Astronomy,  Michigan  State  University,  East Lansing, MI 48824\\
$^{2}$ International Centre for Radio Astronomy Research $--$ Curtin University, GPO Box U1987, Perth, WA 6845, Australia\\
$^{3}$Department of Physics, Box 41051, Science Building, Texas Tech University, Lubbock, TX 79409-1051, USA \\
$^{4}$Eureka Scientific, Inc., 2452 Delmer Street, Suite 100 Oakland, CA 94602, USA\\
$^{5}$ Space Telescope Science Institute, 3700 San Martin Dr, Baltimore, MD 21218}

\date{Accepted XXX. Received YYY; in original form ZZZ}

\pubyear{2018}

\begin{document}
\label{firstpage}
\pagerange{\pageref{firstpage}--\pageref{lastpage}}
\maketitle

\begin{abstract}

    RZ2109 is the first of several extragalactic globular clusters shown to host an ultraluminous X-ray source. RZ2109 is particularly notable because optical spectroscopy shows it has broad, luminous [OIII]$\lambdaup\lambdaup$4959,5007 emission, while also having no detectable hydrogen emission. 
     The X-ray and optical characteristics of the source in RZ2109 make it a good candidate for being a stellar mass black hole accreting from a white dwarf donor (i.e. an ultracompact black hole X-ray binary).
     In this paper we present optical spectroscopic monitoring of the [OIII]5007 emission line from 2007 to 2018. We find that the flux of the emission line is significantly lower in recent observations from 2016-2018 than it was in earlier observations in 2007-2011. We also explore the behaviour of the emission line shape over time. Both the core and the wings of the emission line decline over time, with some evidence that the core declines more rapidly than the wings. However, the most recent observations (in 2019) unexpectedly show the emission line core re-brightening.  
\end{abstract}




\section{Introduction}
RZ2109 is a globular cluster associated with the elliptical galaxy NGC 4472, and is host to a stellar mass black hole candidate. The X-ray source,  XMMUJ122939.7+075333, associated with RZ2109 is ultraluminous ($\sim 10^{39}$ erg s$^{-1}$) and highly variable, with the count rate dropping by a factor of seven in only a few hours \citep{2007Natur.445..183M}. Optical spectroscopy obtained by \citet{2007ApJ...669L..69Z} revealed a very broad and luminous [OIII]$\lambdaup\lambdaup$4959,5007 emission lines, with a velocity width of around 2000 km s$^{-1}$ \citep{zepf08}. The oxygen emission is also variable \citep{2011ApJ...739...95S}. RZ2109 is the only globular cluster associated with NGC 4472 to show [OIII]5007 emission, which places constraints on geometric beaming of the system. If the large X-ray luminosity were produced by strong beaming effects, it would be likely that there would be several sources with strong [O III] whose X-ray emission was beamed away from our line of sight \citep[see][for more detail]{peacock2012a}. 

Not only is the [OIII] emission in RZ2109 broad, but \citet{2012ApJ...759..126P} found that the oxygen emission line is spatially resolved by HST, which implies that there is an oxygen nebula within the cluster with a size scale of $\sim$5$\pm$2~pc. Such a large size for the nebula negates a previous hypothesis that this oxygen emission  is due to ionised nova ejecta that is serendipitous in a globular cluster with a bright X-ray source as suggested by \citet{2012MNRAS.423.1144R}, as the nova ejecta shell required to produce only [OIII] emission must be at least an order of magnitude smaller. All evidence points to the [OIII] emission being associated with an outflow powered by the ultraluminous X-ray source. 

\citet{2014ApJ...785..147S} modelled the emission line nebula and found that the material is hydrogen-depleted, and that the emission from the material is consistent with accretion-powered outflow driven by a white dwarf donor being accreted onto a black hole, making XMMUJ122939.7+075333 (hereafter RZ2109) one of the very few candidate ultracompact X-ray binary (UCXB) systems in a globular cluster. UCXBs are hydrogen deficient systems with a white dwarf donor and either neutron star or black hole accretor,  and typically have very short orbital periods of 80 minutes or less \cite[see][]{2012A&A...537A.104V, Heinke13}. 

In addition to being a strong black hole UCXB candidate, the RZ2109 source is also one of the few known stellar mass black hole candidates in a globular cluster \citep{2012ApJ...759..126P,2018arXiv180601848D}, and possibly the brightest one. Recent theoretical work has shown that GCs are likely to host BHs, and even suggests that black holes drive the whole dynamical evolution of clusters,  \citep[e.g.][]{2019ApJ...871...38K}, so it is not surprising studies are starting to find them, even if their signatures are challenging to observe. Currently there are a handful of accreting black hole candidates in Milky Way globular clusters \citep{2012Natur.490...71S,2013ApJ...777...69C,2015MNRAS.453.3918M, 2018ApJ...855...55S}, and there is one dynamically-confirmed black hole candidate in which there is no evidence for accretion \citep{2018MNRAS.475L..15G}. There are a handful of other black hole candidates in extragalactic globular clusters, including globular clusters in NGC 1399, NGC 4472 and NGC 4649  \citep{2010ApJ...721..323S,irwin2010, 2011MNRAS.410.1655M, 2012ApJ...760..135R, 2019MNRAS.485.1694D}.  Unlike the Milky Way sources, this extragalactic population of globular cluster black hole candidates are all ultraluminous X-ray sources (ULXs), and are likely exhibiting a very different accretion regime than their very much closer analogues. Some ULXs were found to exhibit coherent pulsations \citep[e.g.][]{2014Natur.514..202B}, implying that the compact objects in the binary are neutron stars.  Many models for neutron star ULXs also involve extremely high magnetic fields \citep{2018NatAs...2..312B}. These may be plausible for recently formed neutron stars in star forming regions in the field, but implausible for the old stellar populations of globular clusters. The best physical explanations for these systems requires significant beaming \citep{2017MNRAS.468L..59K}, which has been ruled out for RZ2109 \citep{peacock2012a}, see also \citet{2019MNRAS.485.1694D}.

Of the extragalactic systems,  RZ2109  is the best studied, having been monitored long-term in both X-ray and optical. Studying this system can address questions about the evolution of black holes in globular clusters, and shed light on the nature of black holes in globular clusters, as simulations predict their presence \citep[e.g.][]{2013ApJ...763L..15M, 2015ApJ...800....9M, 2016PhRvD..93h4029R}. The nature of black holes in globular clusters has become increasingly relevant in light of the LIGO discoveries, as globular clusters are one possible source for the progenitors of the merging black hole binaries detected by LIGO \citep{abbottb}. In fact, recent simulations suggest that multiple generations of black hole binaries can form in globular clusters \citep{2019arXiv190610260R}. 

In this work, we present new optical spectroscopy of RZ2109 from 2011 to 2018 and add to the \citet{2011ApJ...739...95S} study of the optical variability beginning in 2007. RZ2109 is also highly variable in X-ray \citep{2007Natur.445..183M, 2008MNRAS.386.2075S}. The X-ray source, which is likely to drive the ionisation of the oxygen nebula \citep{2012ApJ...759..126P} has been monitored long-term in X-ray, from 2000-2016 \citep{2018arXiv180601848D}. We compare the variability of the flux of the oxygen line to the X-ray variability from \citet{2018arXiv180601848D} to search for a link between the variability in both wavelengths. Section \ref{data} describes the data and analysis. The  implications of these measurements on the size scale of the system in Section \ref{results}, and Section \ref{conclusions} discusses the impact of these results. 

\section{Optical Data and Analysis}
\label{data}
Optical spectra have been obtained for RZ2109 since 2007. Previous observations have been reduced and presented in \citet{2011ApJ...739...95S}: it was observed with Keck in 2007 \citep{zepf08}, on the William Herschel Telescope (WHT) in 2008, on the Southern Astrophysical Research Telescope (SOAR) early in 2009, and on the Gemini South Telescope later in 2009.  In this paper we analyse new data starting from 2011 until 2019 to compare to previous observations and measurements. The newest observations were taken with Gemini South/Gemini Multi-Object Spectrograph (GMOS) and SOAR/Goodman High Throughput Spectrograph(GHTS), and are described in more detail below. The data were reduced using IRAF \citep{1986SPIE..627..733T,1993ASPC...52..173T}. 

\subsection{Gemini}
Data were taken using GMOS on Gemini South \citep{2004PASP..116..425H} under observing program GS-2011A-Q-41. Data from both programs span a wavelength range of 4134\AA\hspace{0.1cm} to 5765\AA, with a B1200 grating, resolution R=3744\footnote{\url{https://www.gemini.edu/sciops/instruments/gmos/spectroscopy-overview/gratings}}, and a 1.0\arcsec\hspace{0.1cm} slit. Data for GS-2011A-Q-41 were taken on UT 2011-05-01, 2011-05-02, 2011-05-04, and 2011-05-05. 
RZ2109 was observed on Gemini South in 2015 (GS-2015A-Q-1), however, the 2015 data suffered significantly from defects on the detector which severely impacted the utility of these observations.

\subsection{SOAR}
RZ2109 was observed on the 4.1 m Southern Astrophysical Research Telescope (SOAR) using the GHTS \citep{2004SPIE.5492..331C} on UT 2012-03-14, 2016-03-14, 2018-03-14, 2018-03-15, 2019-04-06 and 2019-04-07. The observations were taken using the 0.95\arcsec\hspace{0.1cm}longslit (0.95\arcsec\hspace{0.1cm}in 2018 and 2019), with the SYZY 930 grating. The resolution is R $\sim$ 1500. It was also observed on 28-04-2019 and 02-05-2019 with a higher resolution (1200 l/mm) grating and 0.95\arcsec\hspace{0.1cm}longslit.   

\subsection{Equivalent Width Measurement}
\label{line}
We develop a method in python to measure the equivalent width of the [OIII] emission of the un-normalised spectra in the following manner: we select spectral bandpasses with wavelengths longer and shorter than that of the [OIII] emission and which are relatively featureless and hence good indicators of the cluster continuum.  An average value is drawn from two regions, a lower region of 4800-4875\AA\hspace{0.1cm}, and an upper region of 5100-5175\AA\hspace{0.1cm}\footnote{\url{https://github.com/kcdage/equivalent_width}}. Out of these two regions, we  randomly select 25 \AA\hspace{0.1cm}wide regions over which to average.   Similar to \citet{2011ApJ...739...95S}, we measure the [OIII] emission between 4964\AA\hspace{0.1cm} and 5058\AA\hspace{0.1cm}(due to the breadth of the emission lines - $\sim$1475 km s$^{-1}$, \citealt{2007ApJ...669L..69Z}). 

\begin{figure}
\includegraphics[width=9cm]{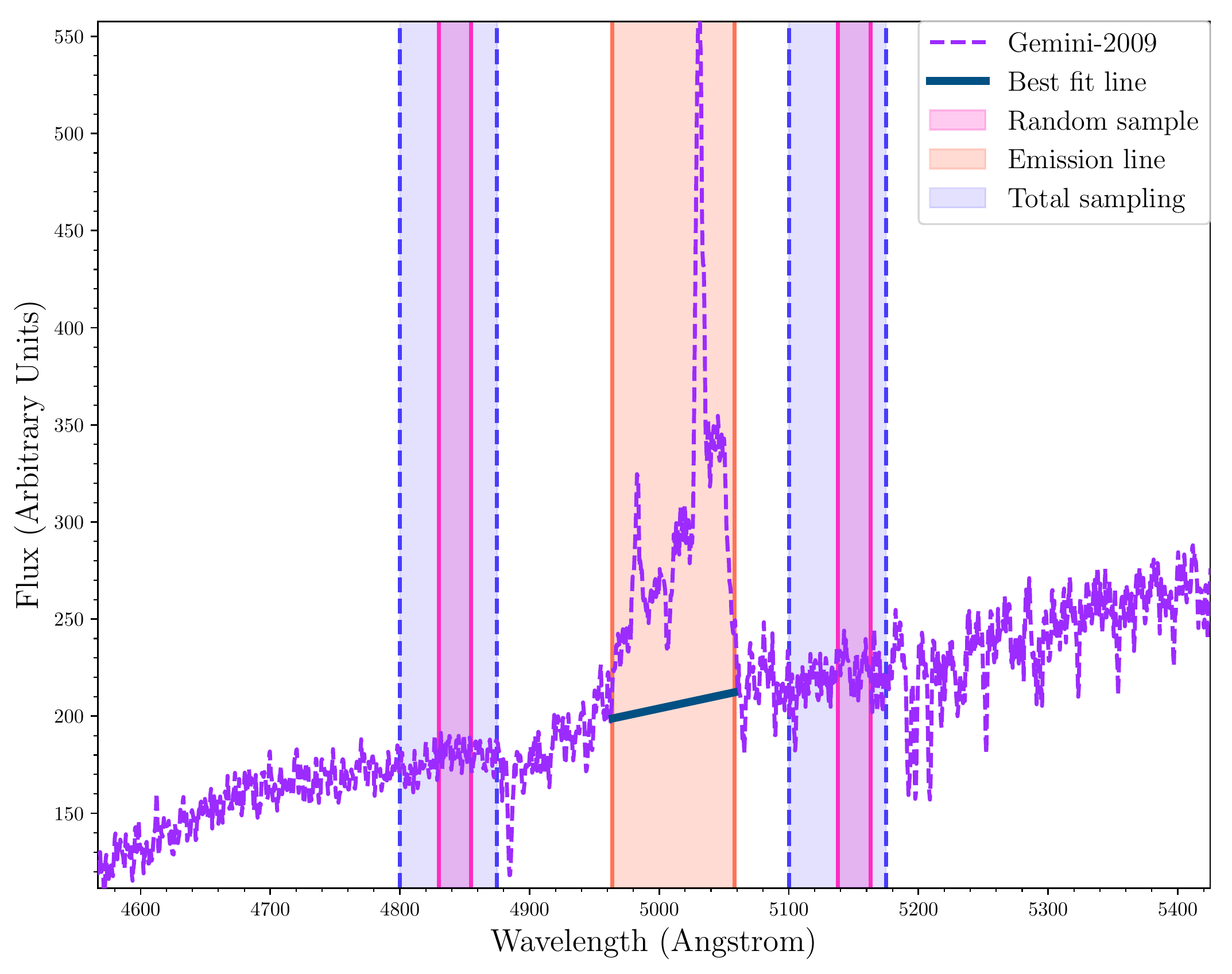}
  \caption{Un-normalised Gemini spectrum from 2009. The purple shaded regions show one example of the 25\AA\hspace{0.1cm} wide regions sampled out of larger regions (blue) that are averaged to approximate the continuum values at the start and end of the emission region. The orange shaded area shows the emission region we are  measuring over from  4964\AA\hspace{0.1cm} to 5058\AA\hspace{0.1cm}. The blue line is the continuum fit  in the region of interest.}

  \label{fig:fitexample}
\end{figure}

We adopt the average value in the bluer range as the continuum value at 4964\AA\hspace{0.1cm}, and the average value in the redder range as the continuum value at 5058\AA\hspace{0.1cm}, and fit a straight line across the broad emission region. The very much lower S/N 2009 and 2012 SOAR spectra were smoothed with a 1D box filter\footnote{\url{http://docs.astropy.org/en/stable/api/astropy.convolution.Box1DKernel.html}}. 
  Figure \ref{fig:fitexample} shows one example for the 2009 Gemini spectrum. We calculate the equivalent width, and to estimate uncertainties, repeat this process 100 times, randomly drawing different regions to calculate the continuum values from. The final reported equivalent width is the average of all these trials, and the error comes from the standard deviation.



The equivalent width of RZ2109  has been measured in \citet{2011ApJ...739...95S}  from observations spanning from 2007-2012, taken with Keck, WHT, SOAR, and Gemini. We also confirm our measurements by remeasuring a subsample of the spectra from \citet{2011ApJ...739...95S}. The data are presented in Table \ref{newfluxtable}.


\begin{table}
\centering
\caption{[OIII] Equivalent Width (Angstrom).}
\label{newfluxtable}
\begin{tabular}{|l|l|l|l|}
\hline
Date       & Instrument & Equivalent Width & Error \\ \hline
2007-12 & Keck  & 31.6 & 1.6\\
2008-01 & WHT & 29.3& 3.2 \\
2009-02 & SOAR & 25.3  & 7.3 \\
2009-03 & Gemini &  32.7 &  0.7\\
2011-05    & Gemini     & 26.0             & 0.6   \\
2012-03 & SOAR & 24.4		& 3.7  \\
2016-03-14 & SOAR       & 18.9             & 0.9  \\ 
2018-03-14, 2018-03-15 & SOAR       & 12.9            & 0.9   \\ 
2019-04-06,2019-04-07& SOAR       & 16.5             & 1.1  \\ \hline
\end{tabular}
\end{table}
\subsection{Normalisation of the Spectra}
\begin{figure}

\includegraphics[width=9.5cm]{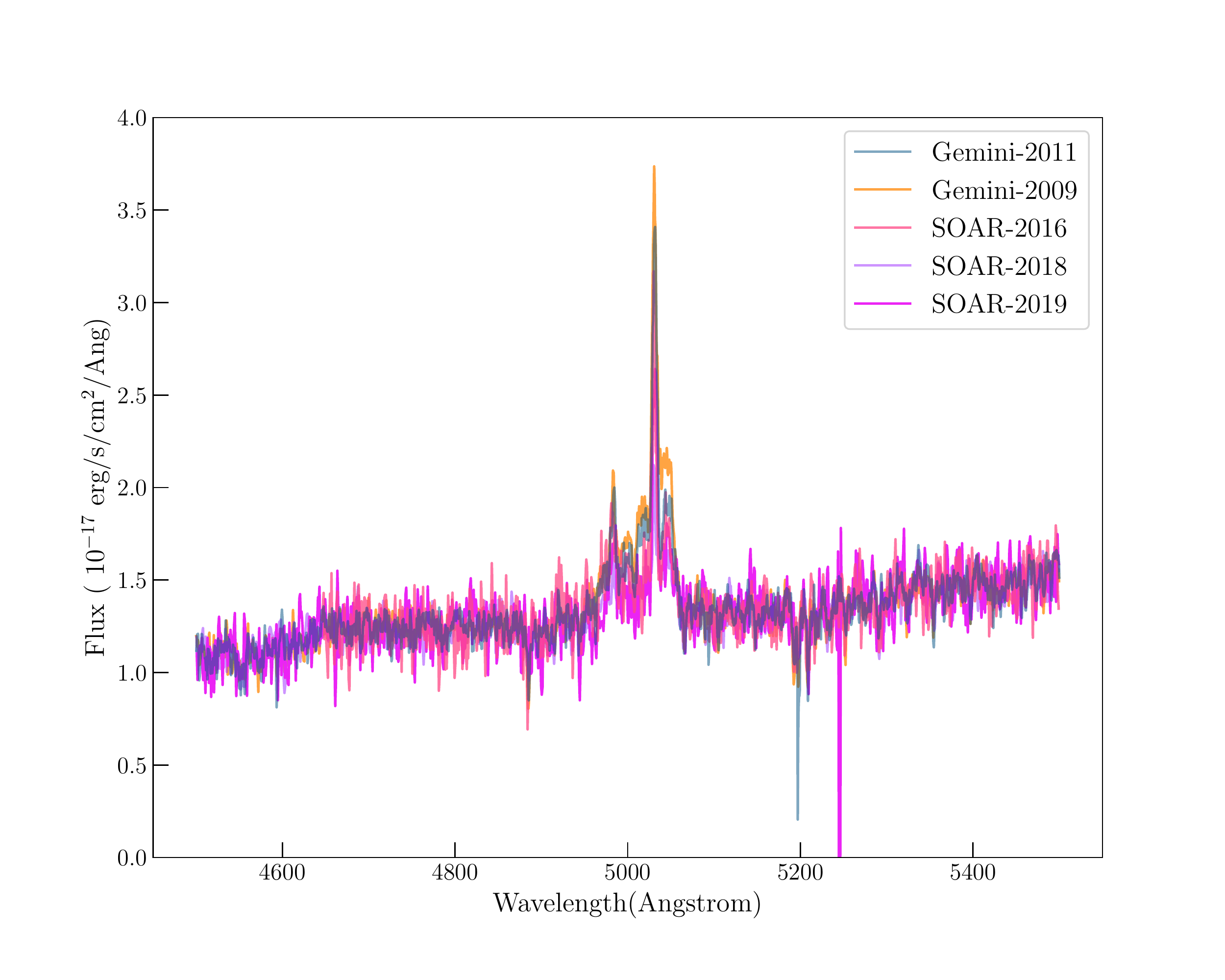}

\caption{Normalised spectra from observations spanning 2007-2018. Observations are from the W.M. Keck Observatory (December 2007, [dark blue]) Gemini South (March 2009, [yellow], May 2011, [light blue]), and SOAR (March 2016, [light pink],  March 2018, [light purple], April 2019 [dark purple]).} 
\label{fig:normspec_ox}
\end{figure}

 All further analysis in this paper is based off of the higher resolution Keck, SOAR and Gemini observations.  The spectral resolution of the 2009, 2011 Gemini spectra are within about 20\% of the 2016, 2018 and 2019 SOAR spectra. We normalise these spectra with respect to the Keck spectrum (which has been flux calibrated) using the following procedure (in python)\footnote{\url{https://github.com/kcdage/spec_slope_fit/}}: we calculate a moving average across 4500\AA\hspace{0.1cm}to 5500\AA\hspace{0.1cm}, but ignores 4940\AA\hspace{0.1cm}-5100\AA\hspace{0.1cm} in order to mask the [OIII] emission\footnote{For the lower wavelength range,we use a lower cutoff of 4646\AA\hspace{0.1cm} instead of 4500\AA\hspace{0.1cm} in the case of the SOAR 2016 data, which had a narrower wavelength range.}. Then we fit a 5th order polynomial to the smoothed continuum. We divide the fitted Keck slope by the SOAR or Gemini slope to find the relative normalisation as a function of wavelength and apply it to each spectrum. The normalised spectra are presented  in Figure \ref{fig:normspec_ox}. 

\section{X-Ray Data and Analysis}
\label{xray}
RZ2109 was observed by \textit{Chandra} for 30~ks on  2019-04-17 (ObsID 21647), and was followed up by a series of shorter \textit{Swift-XRT} observations triggered after the observed rise of the optical luminosity.

\subsection{Chandra}

RZ2109 was detected in the latest \textit{Chandra} observation, with 7 source counts against a background of 2 counts detected in a 3.5" radius region in the 0.3-10keV energy range, making this a 3 $\sigma$ detection \citep{1986ApJ...303..336G}. Similar to \citet{2018arXiv180601848D}, the source was extracted with CIAO version 4.10 using the \textsc{dmextract} tool \citep{2006SPIE.6270E..1VF} and binned by counts of 1. We fit the resultant spectrum with \textsc{xspec} \citep{1996ASPC..101...17A}, using Cash statistics \citep{1979ApJ...228..939C} to fit, and Pearson Chi-Squared as the test statistic, with solar abundances from \citet{2000ApJ...542..914W}.   

 While RZ2109 is typically best-fit by an absorbed two component disk plus power-law model \citep{2018arXiv180601848D}, given the poor statistics and low number of counts, an absorbed power-law model (\texttt{tbabs*pegpwrlw}, with the hydrogen absorption column fixed to 1.60\footnote{\url{http://cxc.harvard.edu/toolkit/colden.jsp}} $\times 10^{20}$cm$^{-1}$) was the more appropriate choice.  We caution that it is not justifiable to interpret any best fit spectral parameters physically, due to low count numbers, and we fit solely to estimate the X-ray luminosity. The best fit flux was 2.2  $^{+6}_{-0.8}$ $\times 10^{-15}$erg cm$^{-2}$ s$^{-1}$ in the 0.5-8 keV energy range. The best fit power-law index is 3.0$^{+5}_{-1}$.  The X-ray luminosity\footnote{Assuming a distance of 16.8 Mpc \citep{macri}.} of this observation in 0.2-10 keV extended energy range calculated using PIMMS\footnote{http://cxc.harvard.edu/toolkit/pimms.jsp} and a power-law index of 3.0 is 2.0 $^{+5.3}_{-0.7}$ $\times 10^{38}$ erg s$^{-1}$.

\subsection{Swift/X-Ray Telescope}

Swift/XRT data on RZ2109 were taken in three epochs in November and December 2007, March 2010, April and May 2019. We reprocessed all Swift/XRT data using xrtpipeline (HEASoft 6.25). The source was not detected in any individual observations, thus we merged observations in each epoch (using Xselect) and investigated source brightness. The source was detected at $\sim$2-sigma significance in 2007, but was not detected in 2010 or 2019. The X-ray luminosities from 0.2-10 keV are presented in table \ref{swift}.

\begin{table}
\caption{X-ray luminosities from 0.2-10 keV measured by Swift/XRT. }
\label{swift}
\begin{tabular}{|l|l|l|}
\hline
Date                                                       & Exposure Length & $L_X$ (0.2-10 keV) \\
& seconds &  $\times 10^{39}$ erg s$^{-1}$ \\ \hline
2007: 11-13, 12-(25, 27)                  & 8327.0              & 1.79 $^{+1.16}_{-0.98}$                                                                         \\ 
2010: 3-(22, 26, 30)                 & 6061.0              & $\leq$ 2.45                                                                                   \\ 
2019: 4-15, 5-(13, 17, 27, 31) & 5800.0              & $\leq$ 1.26                                                                                   \\ \hline
\end{tabular}
\end{table}

\section{Results: Time Behaviour of [OIII] Emission}
\label{results}

Based on the observed changes in the [OIII] emission, we can consider possible implications these changes over time would have for the size scales of the oxygen nebula. In this section, we specifically address what could be causing the changes to the relative strengths of components of the  [OIII]$\lambdaup$5007  emission line, and if we see any link between the change over time in X-ray flux and  [OIII]$\lambdaup$5007 emission flux.

\subsection{Time Evolution of the [OIII] emission}
\label{time}
\begin{figure*}
\includegraphics[width=18cm]{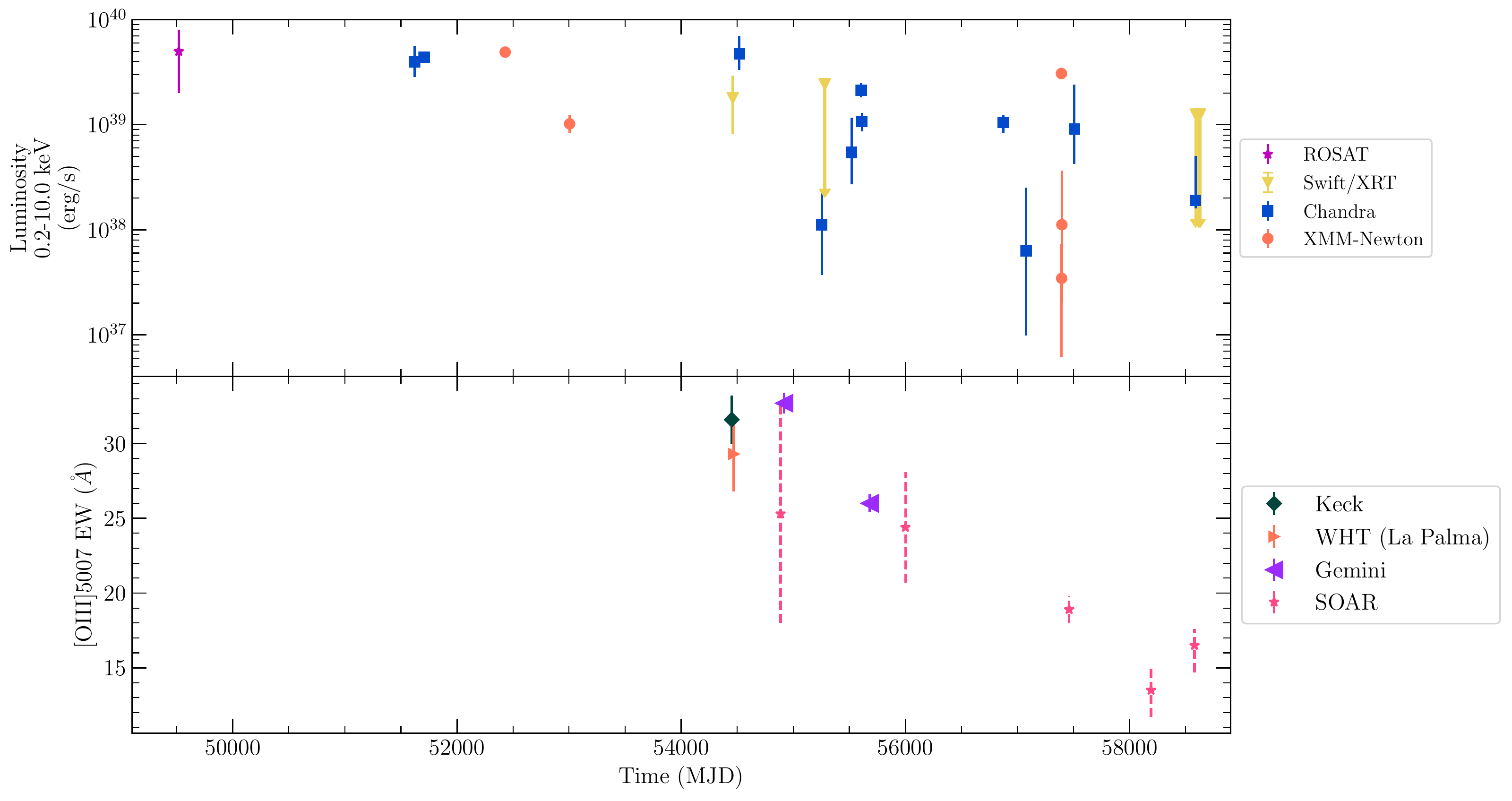}
  \caption{Upper panel: Change in X-ray luminosity over time. Purple star is \textit{ROSAT}, Blue squares are \textit{Chandra} observations, orange circles are taken from \textit{XMM-Newton} observations, yellow triangles are upper limits from \textit{Swift/XRT}. See \citet{2010MNRAS.409L..84M}, \citet{2018arXiv180601848D}, and Section \ref{xray} of this paper for X-ray analysis. Lower panel: Change in L(5007) over time, L(5007) equivalent widths are presented in Table \ref{newfluxtable}. }  

  \label{fig:optvar}
\end{figure*}

We have been monitoring the [OIII]$\lambdaup\lambdaup$4959,5007 emission of RZ2109 since 2007 to examine how the emission has changed over time, and how changes in the optical emission compare to any changes in the X-ray luminosity.
Figure \ref{fig:optvar} plots the full [OIII]$\lambdaup$5007 equivalent widths from \citet{2011ApJ...739...95S} and Table \ref{newfluxtable} to the X-ray luminosities of RZ2109 taken from \citet{2010MNRAS.409L..84M}, \citet{2018arXiv180601848D}, and Table \ref{swift}. 

One of the goals of this work is to use these data to constrain the size of the
[OIII] emitting nebula. From Figure \ref{fig:optvar} it is clear that the
[OIII]$\lambdaup$5007 luminosity decreases by about a factor of two over a timescale of about 3,000 days. Because the light crossing time of the emitting region provides a rough lower limit to the timescale on which the emission can be seen to vary, the data shown in Figure \ref{fig:optvar} place an upper limit on the size of the emitting region to be about four light years across. 

This approach gives an upper limit to the size of the emitting region, but does not provide further information about its size. An alternative approach is to assume the observed decline in the emission line is due to the oxygen nebula expanding and lowering the overall density of material by a factor of two while the central source remains constant. Given this assumption the size scale can be estimated based on the following equations:

\begin{equation}
r_{\textrm{final}}=r_{\textrm{initial}}+(V_{\textrm{expansion}})\times t_{\textrm{decline}}
    \label{eq:2}
\end{equation}

\begin{equation}
r_{\textrm{final}}^3=2\times r_{\textrm{initial}}^3. 
    \label{eq:3}
\end{equation}

Adopting eight years as the time over which the emission is seen to decline and an expansion rate for the emitting region of 1000 km/s (approximately one half the line width), we find that under these assumptions, the current size scale would be roughly 0.04 pc. We can also compare this to the absolute lower limit for the size of the emitting region set by the volume required at the critical density of the [OIII]$\lambdaup$ $\lambdaup$4959,5007 emission line required to produce the observed line luminosity. As also shown in \citep{2011ApJ...739...95S}, this minimum size is a few $10^{-3}$pc. The half-light radius of the emitting region could be limited to greater than $\sim10^{-3}$ pc at the low end and $\sim$2 pc at the high end. 

\subsection{Time Evolution of the Core and Red and Blue Wings of the Emission Line}
\label{wings}

Differential behaviour between the red and blue wings of the observed [OIII] emission can also provide a constraint on the size scale of the nebula.  We will observe changes in the red wing which should lag changes in the blue wing due to the difference in light travel time between the material moving toward us and material moving away. Any such change is convolved with possible changes in the structure of the emitting region, but such a comparison may still provide a useful test of the consistency of  models of the spatial scale of the nebulae.

\begin{figure*}

\includegraphics[width=20cm]{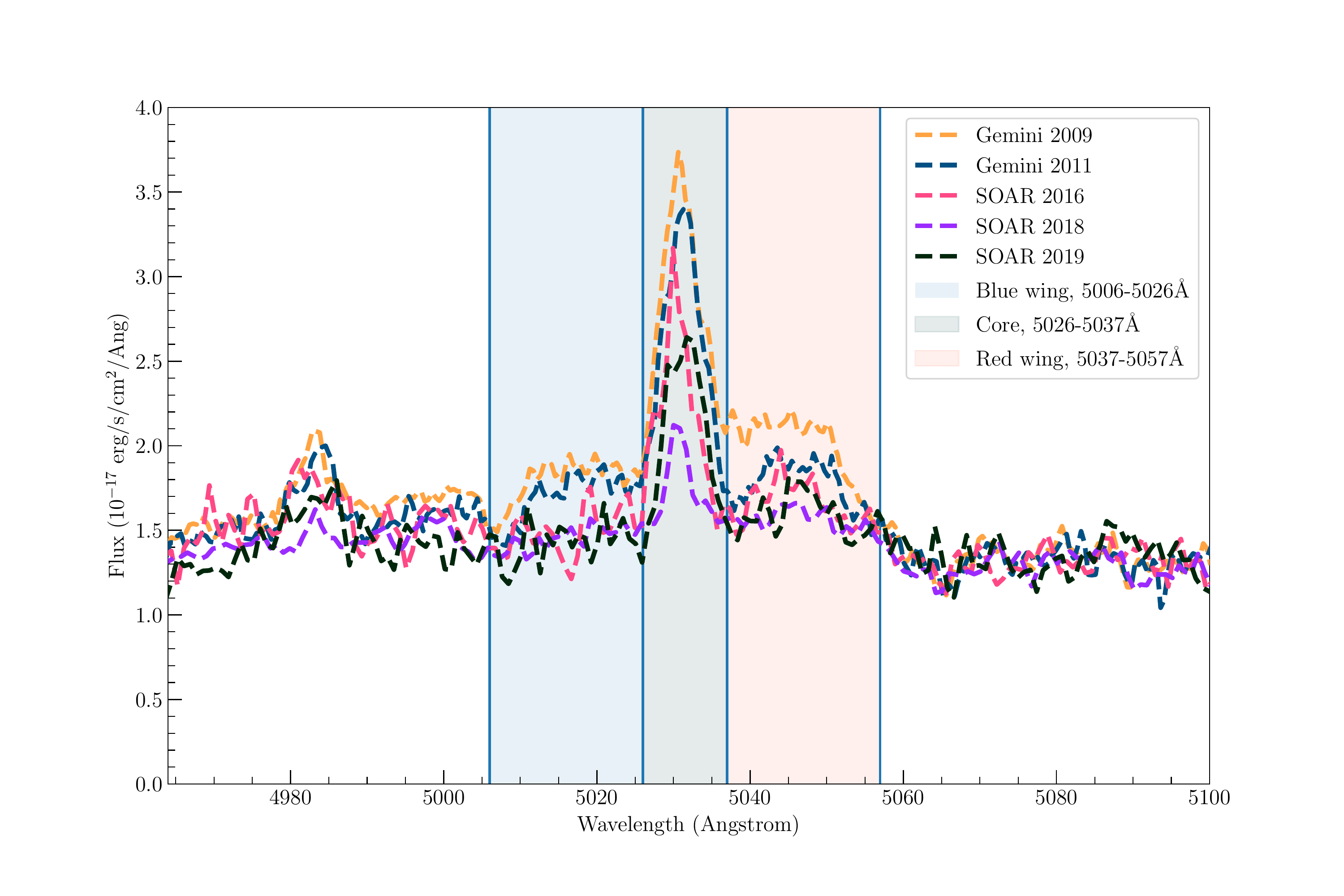}
\caption{Intervals of the  [OIII]$\lambdaup$5007  emission line components, 5006-5026\AA\hspace{0.1cm} for the [OIII] blue wing, 5026-5037\AA\hspace{0.1cm} for the [OIII] core, and 5037-5057\AA\hspace{0.1cm} for the [OIII]  red wing.} 
\label{fig:fluxcore}
\end{figure*}

To estimate this possible lag, we utilise the work of \citet{2014ApJ...785..147S} who modelled the structure of the $\lambdaup$5007  emission region and found that it is best described by a Gaussian core with red and blue shifted wings.  We compute the equivalent widths in the wings and the core of the emission line in the normalised spectra by defining three regions: 5006-5026\AA\hspace{0.1cm}as the blue wing, 5026-5037\AA\hspace{0.1cm}as the core, and 5037-5057\AA\hspace{0.1cm} as the red wing (see Figure \ref{fig:fluxcore}). We calculate the errors  by extending the regions by 2\AA\hspace{0.1cm} on either side, recomputing the equivalent width and taking the difference.  The $\lambdaup$4959 core emission was measured across 4978-4989\AA\hspace{0.1cm}, the same width of region as the $\lambdaup$5007 emission core, but centered around the $\lambdaup$4959 core. The uncertainty was propagated by recomputing the equivalent width for a range of continuum values. 

\begin{figure}
\includegraphics[width=9cm]{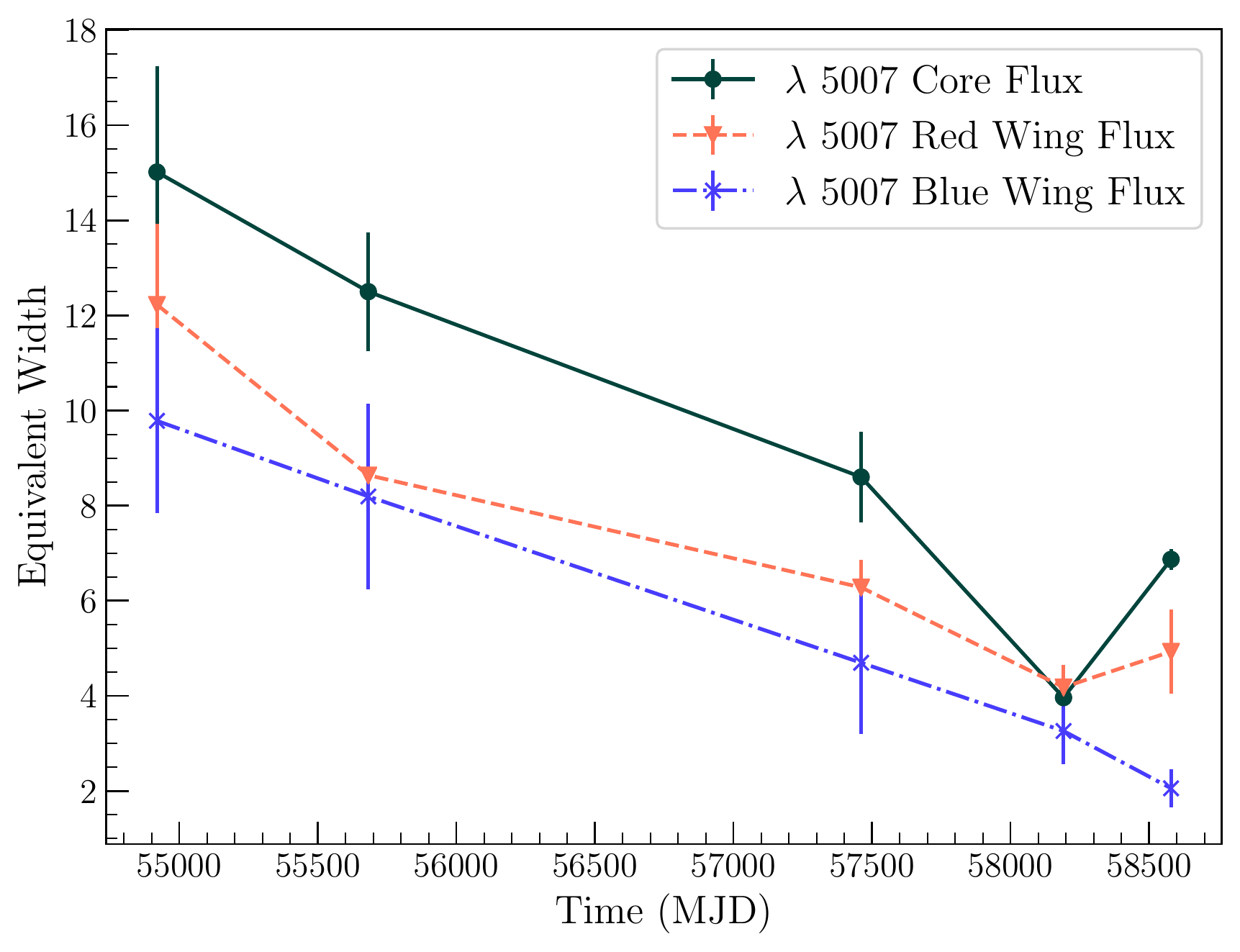}
\caption{Flux in each regime of the emission line over the course of the observations.} 
\label{fig:fluxcontrib}
\end{figure}

\begin{figure}
\includegraphics[width=9cm]{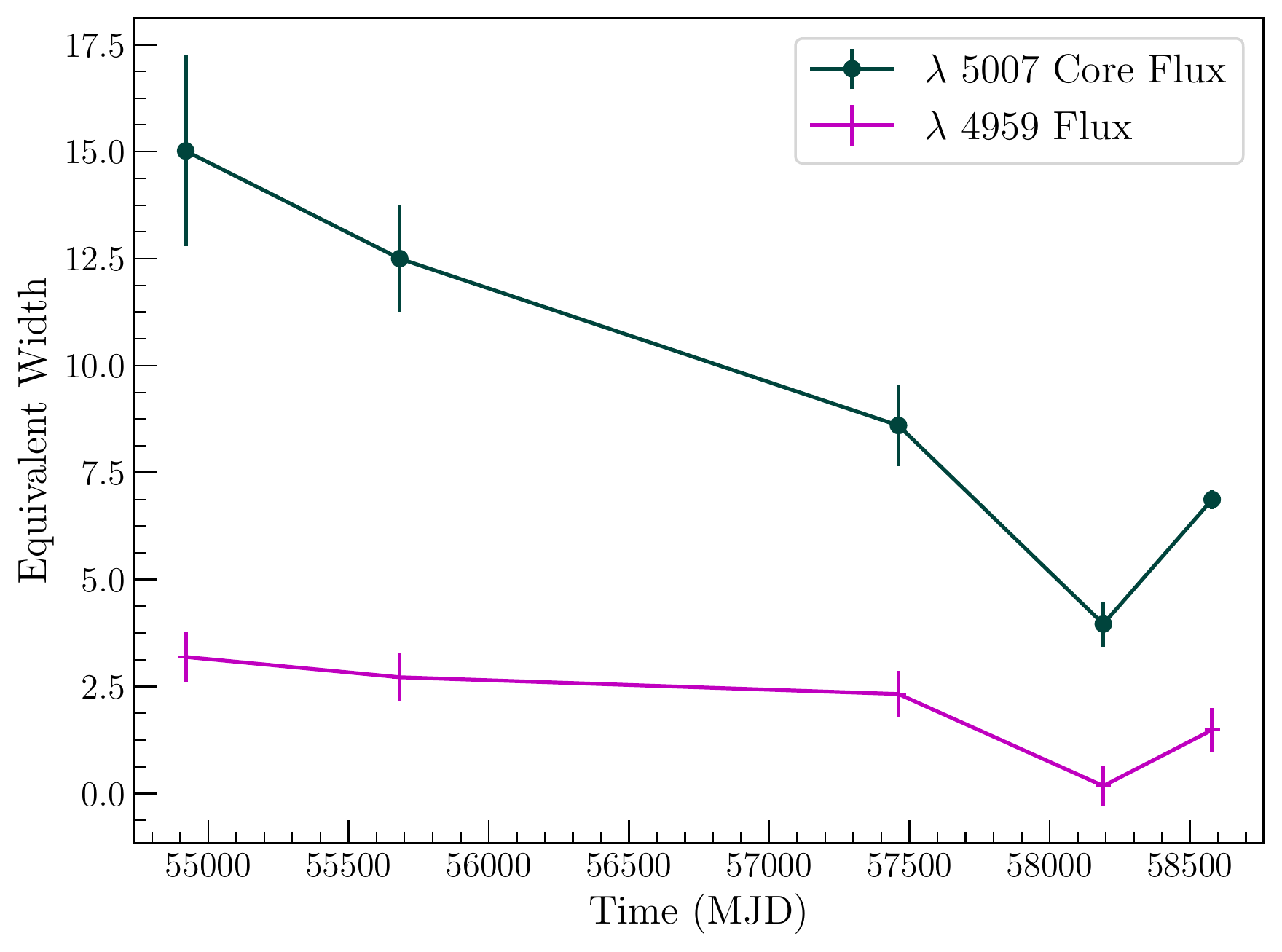}
\caption{Flux in each regime of the $\lambdaup$4959, $\lambdaup$5007 emission cores.} 
\label{fig:fluxcontrib2}
\end{figure}

As can be seen in Figure \ref{fig:fluxcontrib} while all three structures appear to decay with time, they do not appear to do so in concert.
At most epochs, the red emission wing is brighter than the blue wing, but fainter than the core. However, in the past two years, the core continued dropping, then rose, while the red wing seems to have possibly flattened and the blue wing appears to continue to fade. The changes in the core are highly significant compared to the measurement uncertainties, but this is not the case for the wings. Both the $\lambdaup$4959 and $\lambdaup$5007 core emission follow the same trend (Figure \ref{fig:fluxcontrib2}).

As noted above, if the decline of the observed [OIII] flux is due to the central ionising source declining, and the timescale over which it declines is indicative of the size scale of the emitting region, then the red wing of the emission line should lag the behaviour of the blue wing of the emission line, and the time lag should be similar to the overall spatial scale of the emitting region. 

Based on the data presented in Figure \ref{fig:fluxcore}, the red wing may lag the blue wing by around 2500 days ($\sim$ 7 years). This is similar to the size scale derived from the time it takes the overall luminosity to decline by a factor of two. We note that these two measurements are independent - the overall luminosity could decline without any difference in the red and blue wings, or without the red wing being consistent with lagging the blue wing. However, it remains the case that one can construct smaller size scale models where the changes are due to the structure of emitting regions and not due to light travel times.


These implications are a contrast to the spatially resolved HST-STIS spectroscopy analysed by \citealt{2012ApJ...759..126P}. In this work, the HST-STIS spectroscopy appeared to be spatially resolved, with the emitting region having an extent of 5 $\pm$ 2 pc. The variability results appear to suggest a spatial scale on the low end of this very difficult measurement.

\subsection{X-ray Behaviour Versus [OIII]}
Any relation between changes in the X-ray and [OIII] $\lambdaup$5007 luminosities
can be tested by monitoring RZ2109 in both X-rays and [OIII]$\lambdaup\lambdaup$4959,5007 emission. RZ2109 has been observed in both of these wavelengths for more than a decade. In particular, if the optical emitting region is of order a parsec in size, any change observed in the ten or so years we have been observing it must be due to changes in the central X-ray source illuminating the region, as such a large region can not change its structure so quickly. In this case there should be a correlation between the X-ray flux averaged over the light travel time of the system and the [OIII] $\lambdaup$5007 flux.

Figure \ref{fig:optvar} compares the full equivalent widths from \citet{2011ApJ...739...95S} and Table \ref{newfluxtable} to the X-ray luminosities of RZ2109 taken from \citet{2010MNRAS.409L..84M}, \citet{2018arXiv180601848D}, and new data presented in this paper (Table \ref{swift}).

Ideally, we would compare the time changes in the X-ray luminosity to those of the [OIII] $\lambdaup$5007 luminosity to help distinguish between these possibilities for the size of the [OIII] $\lambdaup$5007 emitting nebula. Fundamentally, because the X-ray emitting region is orders of magnitudes smaller than the [OIII] $\lambdaup$ 5007 nebular emitting region for any physically plausible model, the [OIII] $\lambdaup$ 5007 emission should lag changes in the X-ray emission by a time comparable to the light crossing time of the [OIII] $\lambdaup$ 5007 emitting nebula. In order to carry out this test, we need to establish the effective L$_X$ seen by the [OIII] $\lambdaup$ 5007 emitting region on the relevant timescales of months to years. However, determining the effective L$_X$ over longer time periods is made difficult by the widespread short-term variability of RZ2109, so that any X-ray observation may not give a representative value for the L$_X$ at that time. Given this caution it is still notable that the observed L$_X$ in the top panel of Figure \ref{fig:optvar} suggest an overall decline of the L$_X$ of RZ2109.


\subsection{Rise in [OIII] Emission}

\begin{figure}
\includegraphics[width=9.5cm]{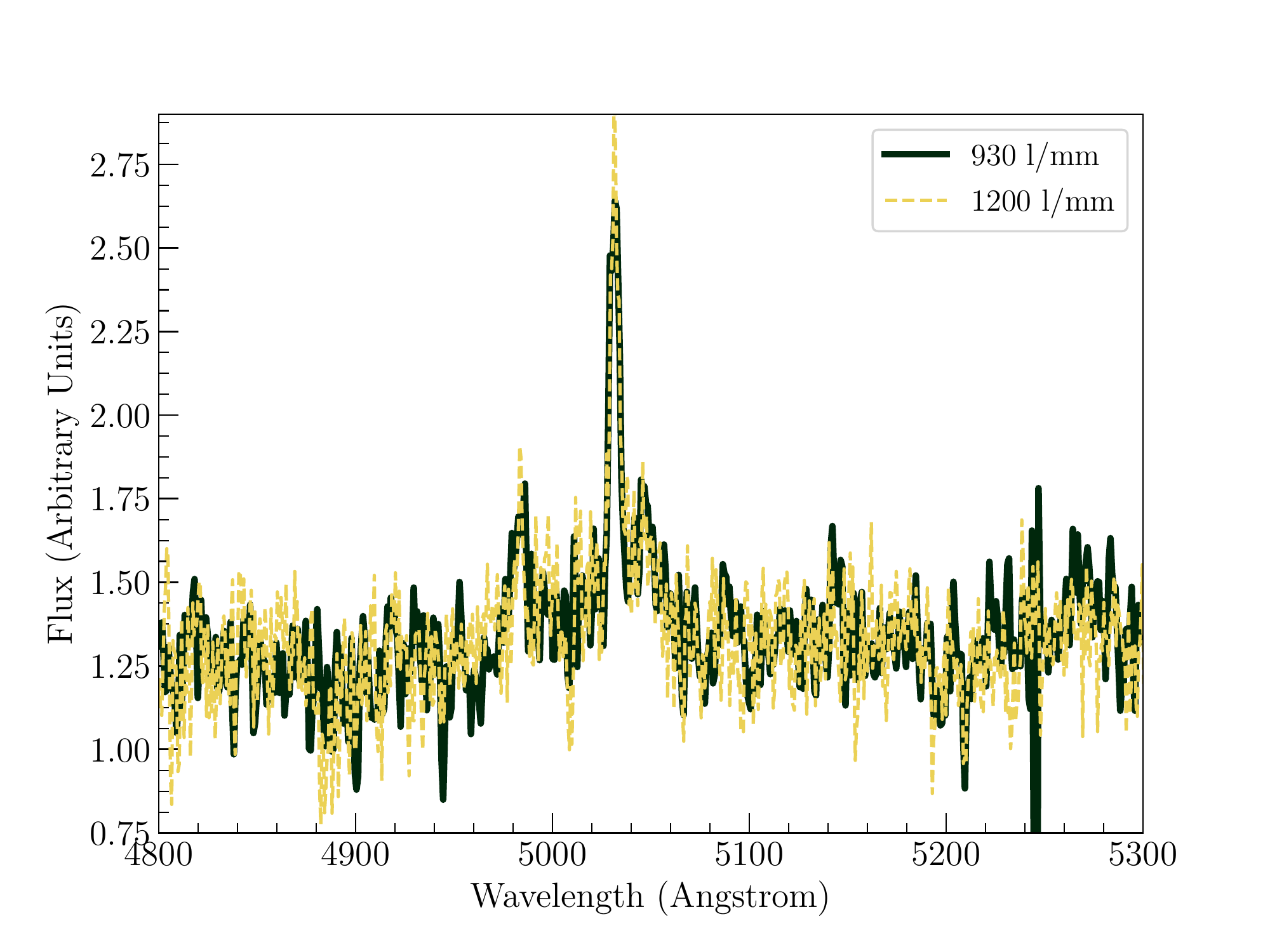}
  \caption{Measurements from early April 2019, and late April 2019 verifying the increase in the [OIII] emission on two different gratings. }  

  \label{fig:newmease}
\end{figure}

\begin{figure}
\includegraphics[width=9.5cm]{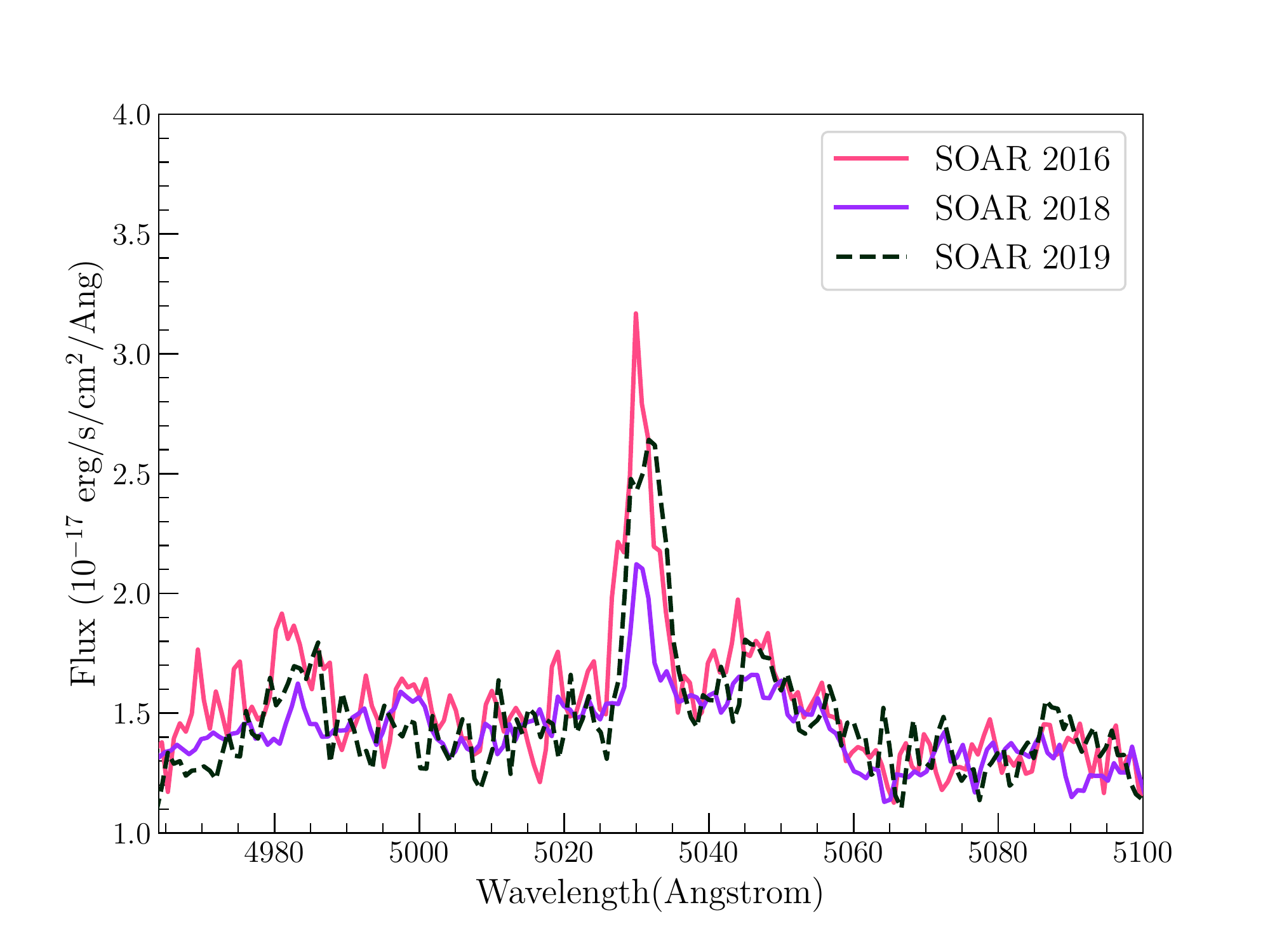}
  \caption{Spectra from 2016-2019 showing decline and subsequent increase. }  

  \label{fig:newmeaseep}
\end{figure}

While the [OIII] emission has been observed in a long-term decline since 2010, in April 2019 the flux in the core was observed to be increasing. Subsequent measurements approximately a month later with a higher resolution grating verify the observed increase (See Figure \ref{fig:newmease}). A close up of the spectrum from 2016-2019 is shown in Figure \ref{fig:newmeaseep} to highlight the decline and rise of the emission line. 

\section{Comparison to Milky Way GC Sources} 

There is one candidate globular cluster BH UCXB candidate among Galactic globular clusters, 47 Tuc X9. This source has $L_X \sim$ a few $\times 10^{33}$ erg s$^{-1}$, but unlike typical compact binaries in of this luminosity, also shows bright radio continuum emission \citep{2015MNRAS.453.3918M}. X-ray timing suggests an orbital period about 28 min, and  the X-ray spectrum shows emission lines consistent with Oxygen VII and VIII and indicates overabundance of oxygen (and possibly carbon) in the system. \citep{2017MNRAS.467.2199B}, but no evidence for optical hydrogen or helium emission/absorption \citep{2018MNRAS.476.1889T}, suggesting a CO white dwarf donor, as inferred for RZ2109. Carbon emission lines were also detected in the FUV spectrum \citep{2008ApJ...683.1006K}. Hence, 47 Tuc X9 appears as reasonable analogue for RZ2109 excepting its lower X-ray luminosity. This can be understood in the context of the evolutionary timescales of UCXBs: RZ2109 is expected to be in a relatively brief period ($\sim 10^5$ yr) of high  mass transfer but will evolve to a fainter, more extended ($\sim 10^7$--$10^8$ yr), longer period system akin to 47 Tuc X9 \citep{2012A&A...537A.104V, church18}. 

We can also consider whether the size scale of the RZ2109 nebula and its implications for the age of the system are consistent with such a scenario. If the size of the RZ2109 nebula is on the order of a parsec, then the system has to be old enough so that the outflowing material can reach this radius. At the observed outflow velocity of about $10^{3}$ km/s, this corresponds to an estimated age of about $10^{3}$ years. Alternatively, if the system is much smaller with a size of $10^{-1} - 10^{-2}$pc, then the age can be correspondingly younger. These are formally lower limits to the age of the system, as the system could have been expanding for longer, and the size scale of the emitting nebula could not be the fullest extent of the ejected material but rather where the density and ionisation parameter are such that it where most of the [OIII] is produced. Therefore, age constraints from the size estimates for the nebula are readily consistent with an age much less than the $10^{5}$ years duration of this evolutionary stage, although they do not absolutely require it.

There is also good reason to believe that the early evolution of the system leads to strong mass ejection. A CO WD donor, as implied by the optical and X-ray data, must have had an initial mass of at least 0.4  $M_\odot$ \citep{2009A&A...507.1575P}, and white dwarfs at or about this mass will require non-conservative mass transfer with strong wind emission to remain in stable mass transfer\citep{2012A&A...537A.104V}. Thus there is both a mechanism to drive oxygen into a nebula and sufficient time for it to reach the size scales we estimate.

\section{Conclusions}
\label{conclusions}
RZ2109 shows very broad [OIII]$\lambdaup\lambdaup$4959,5007 emission, which is also variable. We have been monitoring it with multiple telescopes since 2007 and find that the emission has declined over the last nine years of monitoring. 

\citet{2011ApJ...726...34C, 2012MNRAS.424.1268C} model the X-ray and optical behaviour of an intermediate mass black hole tidally disrupting a white dwarf. The decline of the [OIII] luminosity predicted in the model in \citet{2011ApJ...726...34C} happens much more quickly than what we observe here. Similarly, RZ2109 is declining on a much slower timescale than predicted by the model from  
\citet{2012MNRAS.423.1144R} of a serendipitous bright X-ray source and nova shell ejection. It is also  unlikely that these two unique phenomena are serendipitous in the same cluster.

The oxygen emission is likely caused by X-ray ionisation of the oxygen nebula in the cluster, therefore we expect a link between the X-ray variability and oxygen emission. However, due to the large size of the nebula, it is unclear over what timescales a potential link between the X-ray and optical variability would be observed. The similarity of the declines between both the [OIII] core and  the red and blue high velocity wings are an interesting challenge to models of the physical origin.  

While there may be a hint of a correlation between X-ray luminosity and optical flux, future monitoring in X-ray and  optical of RZ2109 could help determine if this really is the case. Future optical monitoring of RZ2109's [OIII] emission line can help address whether the oxygen emission will continue to decline, and how the broad and narrow components of the [OIII] emission vary.

\section*{Acknowledgements}

KCD, SEZ, and MBP acknowledge support from Chandra grant GO4-15089A. 
SEZ and MBP also acknowledge support from the NASA ADAP grant NNX15AI71G.
This research is based on observations obtained at the Southern Astrophysical Research (SOAR) telescope, which is a joint project of the Minist\'erio da Ci\^encia, Tecnologia, e Inova\c c\~ao (MCTI) da Rep\' ublica Federativa do Brasil, the U.S. National Optical Astronomy Observatory (NOAO), the University of North Carolina at Chapel Hill (UNC), and Michigan State University (MSU). This research has made use of the XRT Data Analysis Software (XRTDAS) developed under the responsibility of the ASI Science Data Center (ASDC), Italy. We  also acknowledge use of NASA's Astrophysics Data System and Arxiv.  KCD acknowledges Jamie Kennea and the Swift ToO Team, and thanks James Miller-Jones and ICRAR/Curtin University. JS acknowledges support from the Packard Foundation. The authors thank the referee for thoughtful and insightful comments.

The following software and packages were used for analysis: IRAF,  distributed by the National Optical Astronomy Observatories,
    which are operated by the Association of Universities for Research
    in Astronomy, Inc., under cooperative agreement with the National
    Science Foundation, SAOImage DS9, developed by Smithsonian Astrophysical Observatory,  \textsc{numpy} \citep{2011arXiv1102.1523V}, Palettable\footnote{\url{https://jiffyclub.github.io/palettable/}}, and  \textsc{matplotlib} \citep{2007CSE.....9...90H}. 





\bibliographystyle{mnras}

\bibliography{rz2109_opt}



\bsp	
\label{lastpage}
\end{document}